\documentclass[12pt]{iopart}

\usepackage[margin=1in]{geometry}
\usepackage{graphicx}
\usepackage{harvard}
\usepackage{subfigure}
\usepackage{amssymb}
\usepackage{filecontents}
\usepackage{color}

\providecommand{\goodgap}{\hspace{\subfigcapskip}}
\providecommand{\ket}[1]{\ensuremath{|{#1}\rangle}}

\begin{document}
\title[]{Multiplexing Superconducting Qubit Circuit for Single Microwave Photon Generation}
\author{R E George$^1$, J Senior$^2$, O-P Saira$^2$, S E de Graaf$^3$, T~Lindstr\"{o}m$^3$, J P Pekola$^2$, Yu A Pashkin$^1$}
\address{$^1$ Physics Department, Lancaster University, Lancaster LA1 4YB, United Kingdom}
\address{$^2$ Low Temperature Laboratory, Department of Applied Physics, Aalto University School of Science, P.O. Box 13500, 00076 AALTO, Finland}
\address{$^3$ National Physical Laboratory, Teddington, Hampton Road, Teddington, TW11 0LW, United Kingdom}

\ead{y.pashkin@lancaster.ac.uk}

\begin{abstract}
We report on a device that integrates eight superconducting transmon qubits in $\lambda/4$ superconducting coplanar waveguide resonators fed from a common feedline. Using this multiplexing architecture, each resonator and qubit can be addressed individually thus reducing the required hardware resources and allowing their individual characterisation by spectroscopic methods. The measured device parameters agree with the designed values and the resonators and qubits exhibit excellent coherence properties and strong coupling, with the qubit relaxation rate dominated by the Purcell effect when brought in resonance with the resonator. Our analysis shows that the circuit is suitable for generation of single microwave photons on demand with an efficiency exceeding 80\%.

\end{abstract}


\submitto{\JPCM}
\maketitle

\section{Introduction}

A new paradigm of information processing based on the laws of quantum physics has triggered intensive research into studying physical systems that can be used as the building blocks of a future quantum processor. Remarkable progress towards realizing quantum information processing elements has been achieved by using both natural and artificial atoms as qubits, and in arranging them into more complex circuits \cite{Buluta2011}. Artificial atoms are engineered quantum systems that have a number of advantages in comparison to their natural counterparts. First, they are fabricated using existing well-developed nanofabrication methods of conventional electronics and therefore can be placed at will whilst having controllable custom-designed features. Second, their size is macroscopic ($\gtrsim 1\,\mu$m), which simplifies the task of coupling multiple qubits together into integrated quantum circuits. The large size of artificial atoms results in their large dipole moment which enables strong coupling of individual qubits to electromagnetic field. Third, their energy levels are tunable by external fields, which simplifies control of the quantum states and inter-qubit couplings. As a downside, the undesired coupling to the environment is strong, leading to shorter coherence times of artificial as compared to natural atoms. This requires careful design of the experimental apparatus to protect fragile quantum states.

Superconducting quantum devices containing Josephson junctions can behave like atoms and are primary candidates to being the building blocks of the quantum processor \cite{Clarke2008}. While there has been enormous progress in the field in the past fifteen years or so \cite{Devoret2013}, the superconducting qubit circuits require further optimsation in order to meet the stringent requirements on coherence for large scale quantum information processing \cite{Martinis2015}. Nonetheless, the presently available circuits are already good enough for doing quantum optics and atomic physics experiments on a chip \cite{Schoelkopf2008,You2011}. A key approach to this is coupling a superconducting qubit to a microwave resonator, thus forming a circuit quantum electrodynamics (cQED) architecture \cite{Blais2004,Wallraff2004}, which is a solid-state analogue of the cavity QED approach \cite{Haroche2013} used to study the interaction of natural atoms with photons. In circuit QED, the field confinement produced by the very small mode volume in combination with the macroscopic size of the qubit results in strong qubit-photon coupling, where quantum excitations can transfer between the artificial atom and the resonator and back many times before decay processes become appreciable.


Circuit QED architecture is uniquely suitable for manipulating microwave radiation at the single photon level \cite{Wallraff2004} and for generation of single microwave photons on demand \cite{Houck2007,Bozyigit2011}. An alternative approach to the single microwave photon generation is based on an artificial superconducting atom directly coupled to an open-end transmission line, a 1D analogue of the 3D half-space \cite{Peng2015}. Here we report on a design comprising eight resonators, each housing a superconducting transmon qubit \cite{Koch2007} that exhibits good coherence properties. The design allows for efficient frequency multiplexed testing and assessment of the superconducting qubits and is suitable for single microwave photon generation. This involves time-domain control of the quantum state in the superconducting circuit, which is a well developed technique commonly used in qubit experiments. The key feature of our design is that it allows for multiplexed generation of single photons with an efficiency of $>80\%$ using quantum state manipulation and dynamic tuning of the circuit parameters. This provides the means of coupling solid-state qubits with each other for long-distance communications as well as linking stationary and flying qubits.

\section{$\lambda/4$ resonator design}

\begin{figure}
\begin{center}
   \subfigure[\label{chip_layout}]{\includegraphics[height=2in]{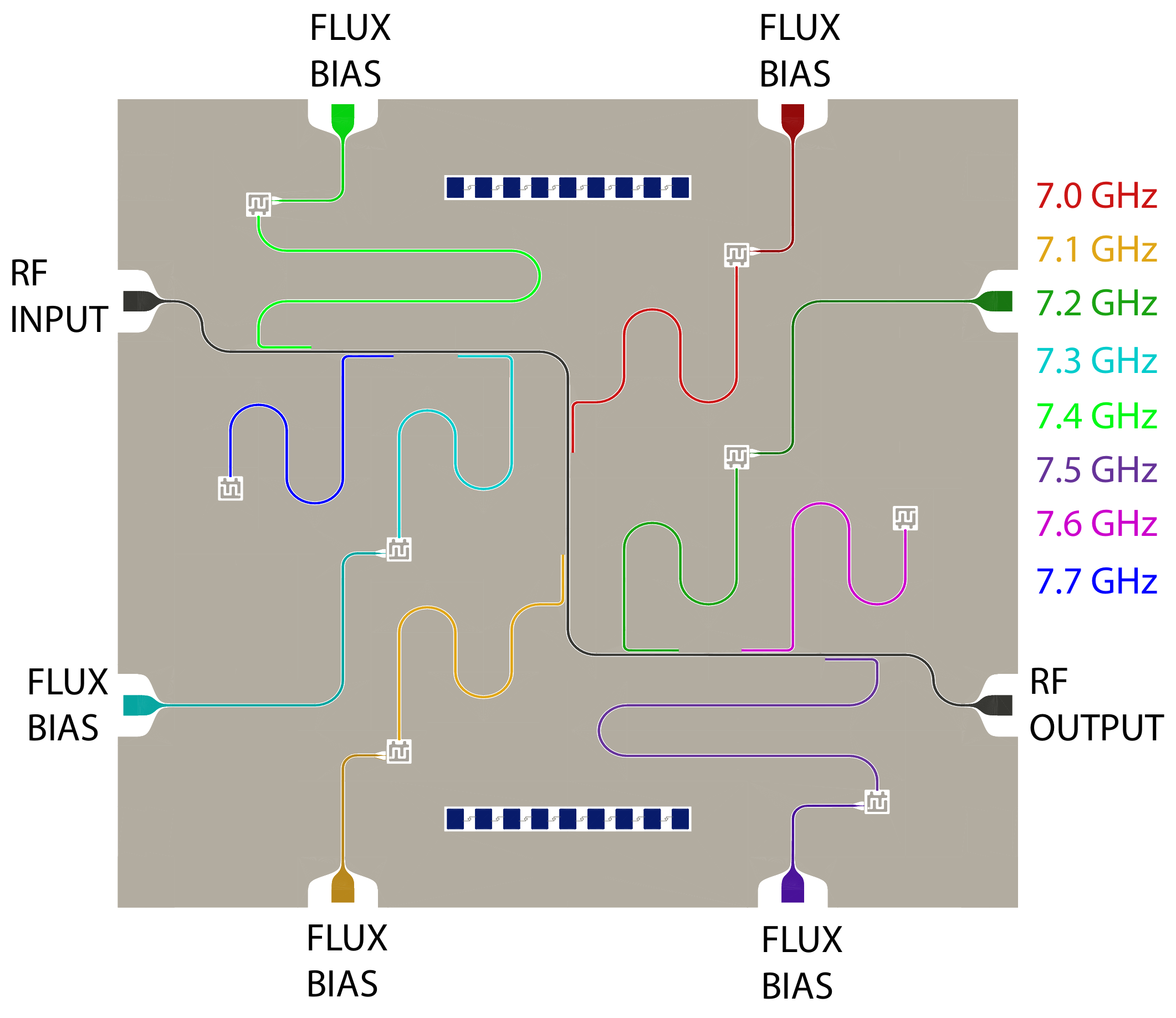}}
    \goodgap
   \subfigure[\label{inductive_coupling}]{\includegraphics[height=2in]{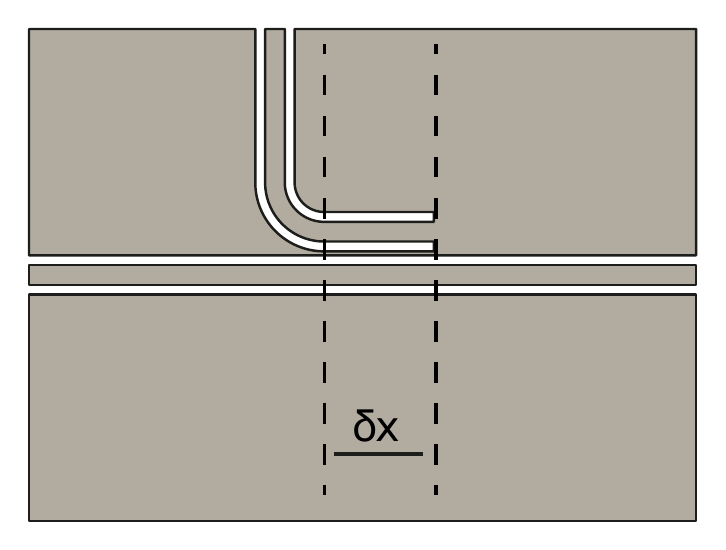}}
    \caption{\label{inductive_coupling_figure}Device layout and realisation of the eight resonator chip. (a) Schematic layout of the chip showing eight resonators coupled to a common feedline, labelled by unloaded resonant frequencies. (b) Inductive coupling section optimised to set external $Q$ factor of resonators. \label{eight_resonator_chip}}
\end{center}
\end{figure}

The resonators are designed as coplanar waveguides in superconducting metal film, having a centreline width W = 20\,$\mu$m and centreline to groundplane spacing of S = 10\,$\mu$m, achieving a characteristic impedance $Z_0 \simeq 50\,\Omega$ on our sapphire substrate. We calculate from the geometry of our transmission lines a capacitance per unit length of $c_r \simeq  153\;\mathrm{pF/m}$, an inductance per unit length $l_r \simeq 402\,\mathrm{nH/m}$, and a phase velocity of $\beta = 53.3\;\mathrm{radians/m/GHz}$ so that a $\lambda/4$ resonator at 7$\,$GHz has a length of 4220\,$\mu$m. The eight $\lambda/4$ resonators are designed with unloaded operating frequencies of 7.0, 7.1, 7.2 $\cdots$ 7.7$\;$GHz by varying their lengths (see figure \ref{chip_layout}). The resonators are inductively coupled to a common feedline running diagonally across the chip (see figure \ref{inductive_coupling}). The coupling strength between the feedline and the resonators is adjusted by a short, $\delta x \sim 400\,\mu$m, section of the resonator parallel to the feedline with centre to centre distance of $44\,\mu$m.


The inductive coupling between the feedline and resonator directly determines the coupling quality factor $Q_c$ of the $\lambda/4$ resonator at the resonator's operating frequency $f_0$. Following Barends \cite{Barends2013} we have that: 

\begin{equation}
Q_c = \frac{\pi}{2 |S_{21}(f_0)|^2}
\end{equation}

\noindent We designed for $Q_c = 5000$, corresponding to an $S_{21}(f_0)$ parameter between the feedline and
resonator of $-35\,$dB, and a resonance linewidth of $\kappa / 2 \pi = 1.4 \, \mathrm{MHz}$. Our design is similar to other frequency multiplexed qubit devices \cite{Jerger2012,Groen2013}

\section{Qubit design}

All the qubits utilised in the experiment are transmons that are anharmonic LC oscillators whose nonlinear inductance is provided by Josephson junctions. The transmon is essentially a charge qubit \cite{Pashkin1999} whose Josephson junction is shunted by a large capacitor that dominates the total capacitance of the device and thus reduces the sensitivity to charge noise \cite{Koch2007}. The qubit is described by the following Hamiltonian:

\begin{equation}
H =  4E_C (\hat{n}-n_g)^2 - E_J \cos \hat{\phi},
\end{equation}

\noindent where $E_C$ and $E_J$ are the charging and Josephson energies, respectively; $\hat{n}$ is the number of Cooper pairs transferred between the superconducting electrodes; $n_g$ is the effective offset charge and $\hat{\phi}$ is the superconducting phase difference between the two electrodes comprising the qubit. The transmon energy level structure is determined entirely by $E_J$ and $E_C$, and it is operated in the regime $E_J \gg E_C$. We designed the qubit to have a maximum $\ket{0} \rightarrow \ket{1}$ transition frequency of $\nu_{01} = 8.5 \; \mathrm{GHz}$, higher than the bare frequencies of the resonator array, by assuming that $h \nu_{01} = \sqrt{8 E_J E_C} - E_C$. For fixed $\nu_{01}$, the ratio $r = E_J / E_C$ influences the qubit dephasing rate through the sensitivity of $\nu_{01}$ to the gate offset charge $n_g$, with large values of $r$ promoting long relaxation times. At the same time, the value of $r$ determines the maximum qubit operation speed through the level anharmonicity $|\nu_{01} - \nu_{12}|$, with faster operation times favouring smaller values of $r < 100$ \cite{Koch2007}. Since in our case the qubit forms part of a single photon source where both parameters are important, we selected a suitable compromise as $r = 35$ which allows for operation times of the order of tens of ns while maintaining long relaxation times ($>1\,\mu$s).

The Josephson energy $E_J$ and charging energy $E_C$ are determined by the device tunnel junction resistance $R_N$ and device self-capacitance $C_\Sigma$, respectively, properties that are under engineering control during fabrication according to the simple relations:

\begin{equation}
R_N = \frac{R_Q}{2} \frac{\sqrt{8r}-1}{r} \frac{\nu_\Delta}{\nu_{01}}
\end{equation}

\begin{equation}
C_\Sigma = \frac{1}{R_Q} \frac{\sqrt{8r} - 1}{8} \frac{1}{\nu_{01}},
\end{equation}

\noindent where the quantum resistance $R_Q = h/4e^2 \simeq 6.46\,\mathrm{k}\Omega$ and $\nu_\Delta = 48.3 \, \mathrm{GHz}$ is related to the superconducting gap $\Delta$ in aluminium used to fabricate the tunnel junctions according to $\Delta = h \nu_\Delta \simeq  200 \, \mu \mathrm{eV}$. Substituting $r = 35$ and $\nu_{01} = 8.50 \, \mathrm{GHz}$ yields $R_N = 8.24 \, \mathrm{k}\Omega$ and $C_\Sigma = 35.8 \, \mathrm{fF}$, corresponding to a Josephson energy $E_J = 78.2 \, \mu \mathrm{eV}$ and charging energy $E_C = 2.23 \, \mu \mathrm{eV}$. It is also useful to express the charging energy $E_C$ as a frequency $\nu_C$ following $ E_C = h \nu_C$, and we find $\nu_C = 540 \, \mathrm{MHz}$.

The final important property of the qubit for our purposes is the state dependent dispersive shift $ \chi / 2 \pi \sim \left. (g/2 \pi)^2 \nu_C \middle/ \Delta_0 (\Delta_0 - \nu_C) \right.$ of the resonator, where $g$ is the coupling strength between the resonator and qubit and $\Delta_0$ is the frequency difference between the bare qubit and resonator states \cite{Bianchetti2009}. When the qubit and resonator are well detuned from one another, the state of the qubit $\ket{0}$ or $\ket{1}$ `pulls' the resonator frequency by $\pm \chi$ from the bare resonator frequency, providing a means to read out the qubit state \cite{Wallraff2005PRL}.

Next we consider determining the coupling capacitance $C_g$ so as to realise a desired dispersive shift $\chi$. We find:

\begin{equation}
\frac{C_g}{C_\Sigma} = \left( \frac{2}{r} \right)^{1/4} \left( \frac{\Delta_0}{\nu^0_\mathrm{rms}} \right) \left( \frac{\chi}{2 \pi \nu_C} \right)^{1/2} \left( 1 - \frac{\nu_C}{\Delta_0} \right)^{1/2},
\end{equation}

\noindent where $\Delta_0 = 8.5 \,\mathrm{GHz} - 7 \, \mathrm{GHz} = 1.5 \, \mathrm{GHz}$ is the qubit-resonator detuning, $\nu_C$ is the charging energy expressed as a frequency (540\,MHz), and $h \nu^0_\mathrm{rms} = e V^0_\mathrm{rms}$ describes the vacuum RMS voltage fluctuations of the resonator. We have that:

\begin{equation}
\nu^0_\mathrm{rms} = \frac{e V^0_\mathrm{rms}}{h} = \left( \frac{1}{2} \frac{\nu_r}{ R_Q \lambda c_r} \right)^{\frac{1}{2}},
\end{equation}

\noindent so that in our $\lambda/4$ resonators, $V^0_\mathrm{rms} = 2.58 \, \mu\mathrm{V}$ and $\nu^0_\mathrm{rms} = 623 \, \mathrm{MHz}$. Substituting the values above determines the desired coupling capacitance as $C_g = 0.09 C_\Sigma \approx 3.2 \, \mathrm{fF}$.


\noindent The total self capacitance $C_{\Sigma}$ is the sum of the tunnel junction capacitance and the geometric capacitance. We estimate the junction capacitance using a typical value of the specific capacitance for Al ultrasmall tunnel junctions of 45\,fF/$\mu$m$^2$ \cite{Haviland1995}, which for our junction area of $100 \times 100\,$nm$^2$ gives $\simeq 0.45\,$fF. Clearly, the junction capacitance contributes only about 1\% to $C_{\Sigma}$ and we can safely neglect it.

\begin{figure}
\begin{center}
\subfigure[]{\includegraphics[height=1.6in]{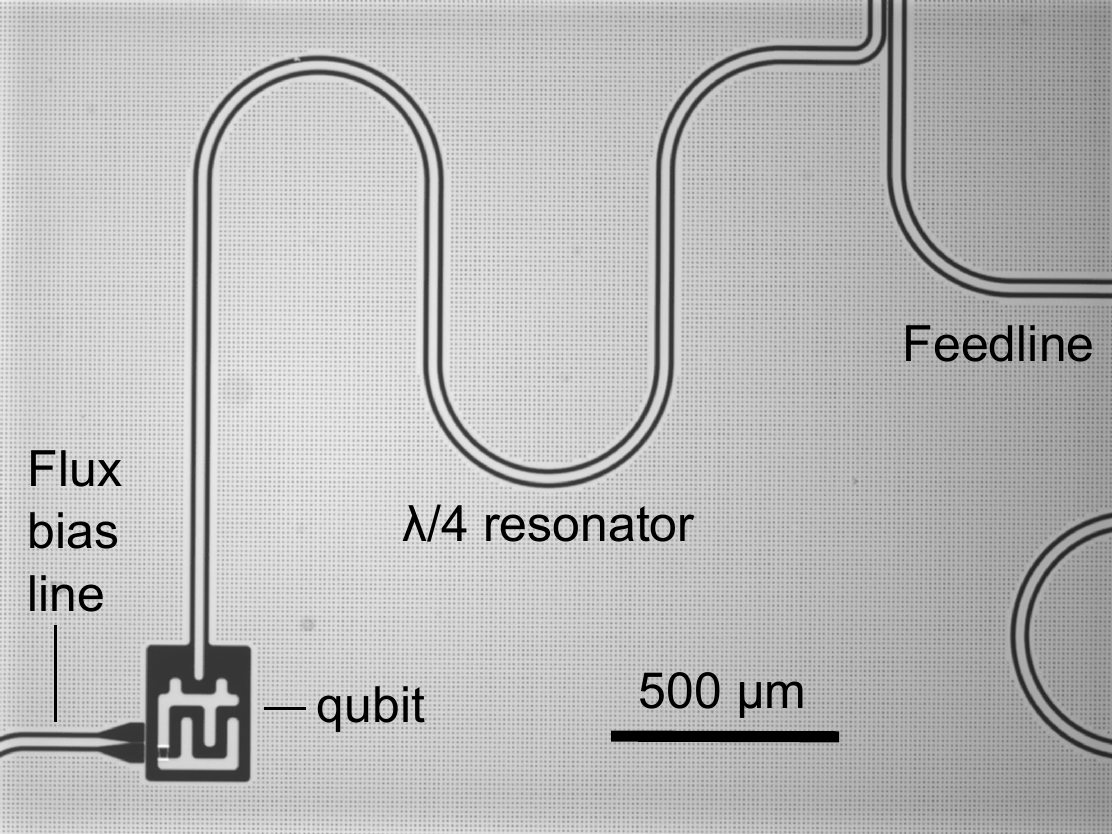}}
\goodgap
\subfigure[]{\includegraphics[height=1.6in]{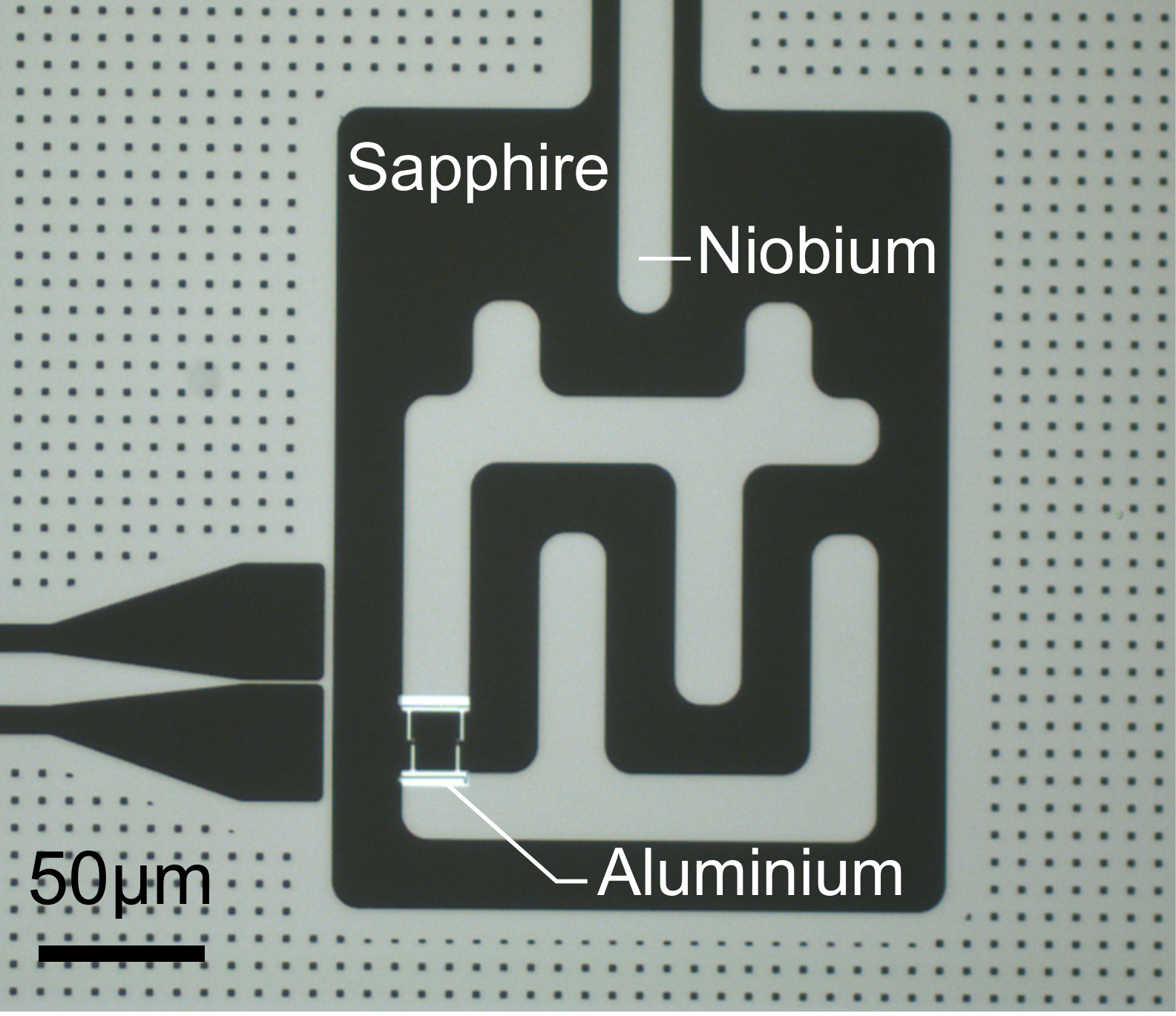}}
\goodgap
\subfigure[\label{squid_loop_image}]{\includegraphics[height=1.6in]{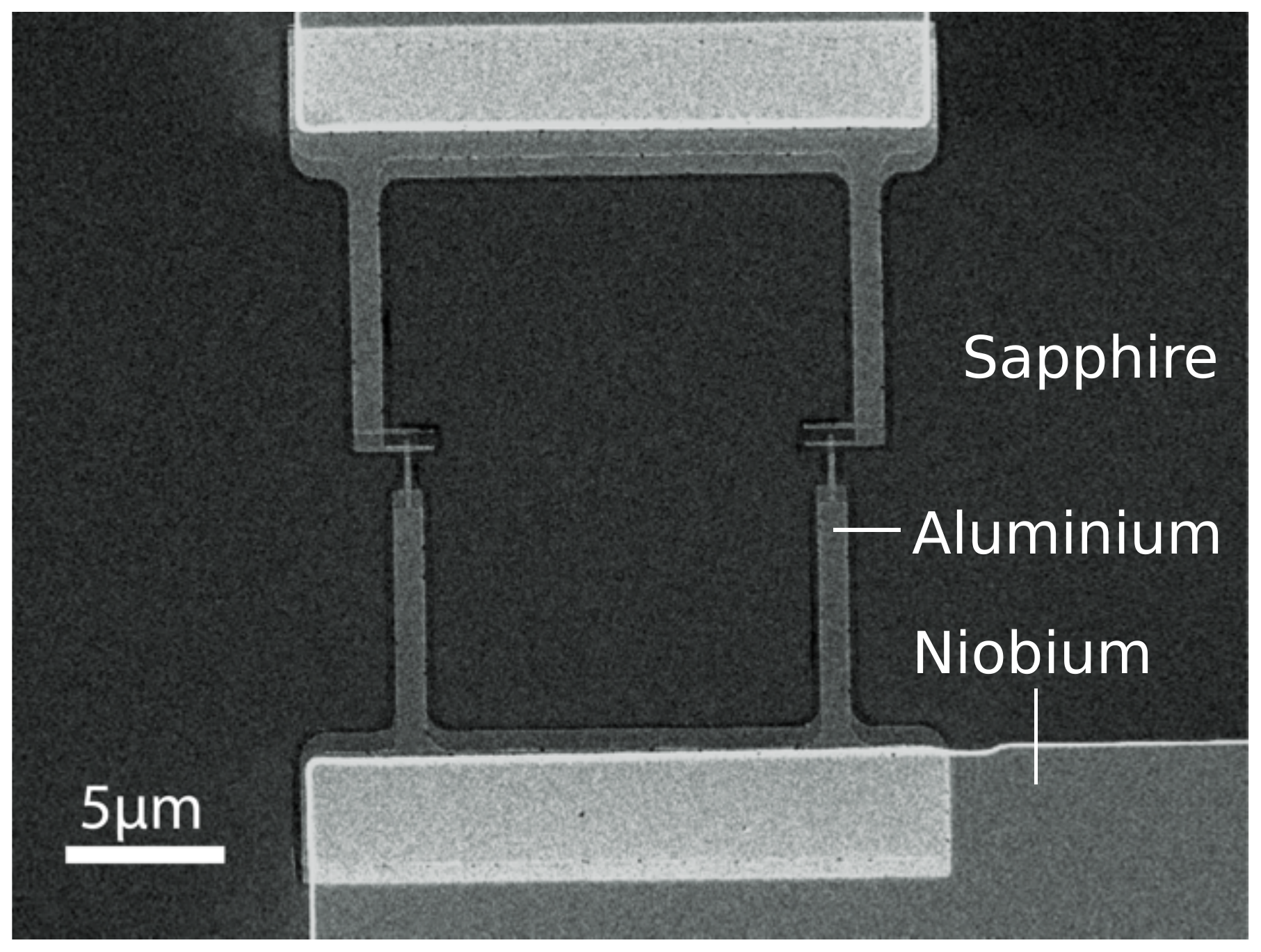}}
\end{center}
\caption{\label{squid_loop}a) A $\lambda/4$ resonator inductively coupled to a feedline and capacitively coupled to a transmon qubit b) Close-up of the transmon qubit c) Scanning electron microscope image of the SQUID loop tunnel junction.}
\end{figure}

The qubit self-capacitance $C_\Sigma$ is then determined through the shunt geometry, whilst the qubit gate capacitance is determined by the separation between the end of the transmission line and the qubit shunt. The device geometry was designed using the COMSOL electromagnetics modelling package to obtain the desired $C_\Sigma$ and $C_g$. To obtain the desired $R_N$ and thus $E_J$, the junction oxidation time and pressure were adjusted to obtain a tunnel resistance of 8.2\,k$\Omega$ measured through the parallel combination of the two nominally $100 \times 100\,$nm$^2$ tunnel junctions in the device SQUID loop. The junction resistance was confirmed by measuring test junctions of the same dimensions as the transmon device, fabricated on the same chip, as shown in figure \ref{squid_loop_image}.

\section{Device fabrication}

The samples were fabricated on a single crystal c-axis oriented sapphire wafer. The fabrication process involved two major steps: patterning of the niobium groundplane, followed by deposition of the transmon structure and tunnel junctions.

{\it Patterning of the groundplane:} The feedline, $\lambda/4$ resonators and qubit shunt capacitors were formed by etching a 100\,nm niobium metal film deposited by DC magnetron sputtering onto the sapphire wafer. The metal coated wafer was spin-coated with AR6200(EL11) resist at 6000\,rpm for 60 seconds, and the resist was baked on a hotplate at 150$^\circ$C for 9 minutes before the resonator and feedline structures were exposed in an e-beam writer to a dose of 350\,$\mu$C/cm$^2$ using a 58\,nA beam current and proximity correction routine. The exposed wafer was developed in AR600-546 developer for 3 minutes before immersing in isopropyl alcohol to halt the development process. The resulting wafer was dried using N$_2$ gas and the resist was reflow baked for 5 minutes at 150$^\circ$C. An SF$_6$+O$_2$ reactive ion etching process was then used to transfer the pattern from the resist into the niobium film. The reactive ion etching chamber was pre-conditioned for 10 minutes before the samples were loaded and etched for 3 minutes 15 seconds, resulting in pattern transfer into the niobium film. The e-beam resist was removed by immersing in AR600-71 solvent for 5 minutes, followed by sonicating in acetone for 1 minute and rinsing in isopropyl alcohol for 1 minute, before  drying in N$_2$ gas.

The patterned wafer was then coated with a protective layer of AZ5214E photoresist before dicing into 8$\times$8mm chips using a hubless resin blade. Following dicing, the protective resist was removed in acetone followed by isopropyl alcohol and N$_2$ drying.

{\it Deposition of transmon tunnel junctions:} The chips were prepared for qubit junction deposition by spin-coating with two layers of MMA(8.5)MAA EL11 resist, at 4000\,rpm for 60 seconds, then baking each layer for 2 minutes at 160$^\circ$C, to create a 1\,$\mu$m film of MMA(8.5)MAA resist, followed by a single layer of 950 PMMA A4 spun at 4000\,rpm for 60 seconds and baked for 10 minutes at 160$^\circ$C. To avoid excessive charging during e-beam exposure, the resist structure was coated with an ``E-spacer 300z'' conductive layer by spinning at 4000\,rpm for 60 seconds. The qubit patterns were exposed using a beam current of 1.1nA on fine structures and 10\,nA on coarse structures. The exposed resist was developed by rinsing in flowing deionised water for 30 seconds to remove the E-spacer layer, followed by immersion in 1:4 vol:vol methyl isobutyl ketone:isopropyl alcohol developer for 20 seconds, and then immersion in a 1:2 vol:vol mixture of methyl glycol:methanol for 20 seconds, followed by immersion in isopropyl alcohol for 10 seconds to halt development, before blowing dry with N$_2$ gas.

The resulting resist structure was used to deposit the qubit junctions in an electron beam evaporation chamber equipped with a tilting sample stage, argon ion milling capability and O$_2$ inlet needle valve. The sample surface was first cleaned by ion milling for 5 minutes using 1\,kV argon ions. A 20\,nm aluminium film was evaporated whilst the stage was tilted at $-17^\circ$. The aluminium was then oxidised in 4\,mbar O$_2$ for 4 minutes before a second 20\,nm aluminium layer was deposited with the stage tilted to +17$^\circ$. The resulting chips were subjected to a lift-off process in warm acetone at 52$^\circ$ for 30 minutes, followed by a cleaning rinse in fresh acetone, isopropanol and drying with N$_2$ gas. The normal state resistance of the parallel combination of two nominally 100$\times$100\,nm$^2$ tunnel junctions was measured and found to be 8.2\,k$\Omega$.

\section{Device characterisation}
\subsection{Measurement setup and resonator characterisation}

\begin{figure}
\begin{center}
\includegraphics[width=0.6\textwidth]{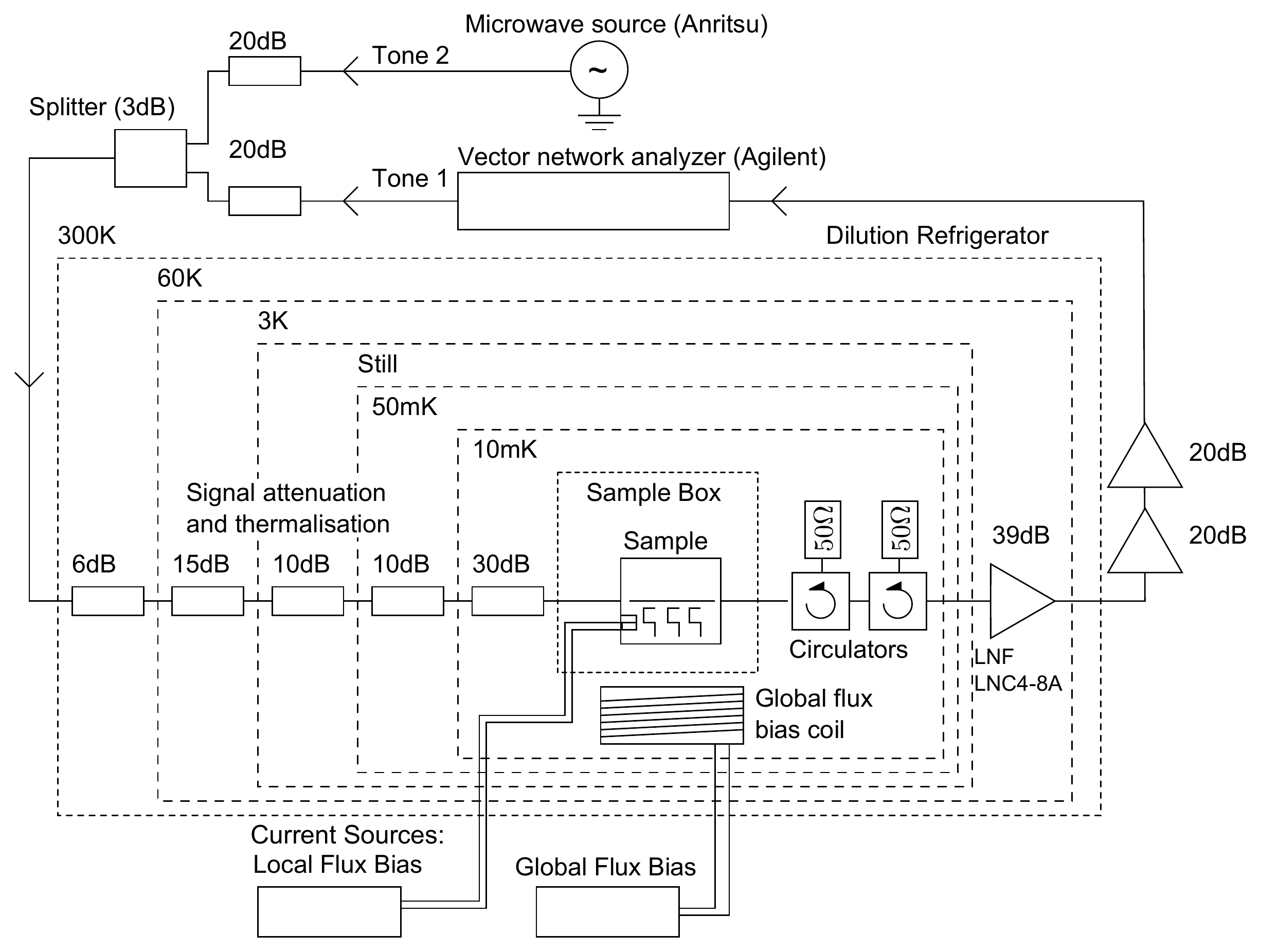}
\end{center}
\caption{\label{measurement_setup}A schematic of the measurement setup showing the microwave attenuation at each stage of the dilution refrigerator.}
\end{figure}

The chip containing superconducting elements was bonded to a printed circuit board and mounted at the mixing chamber plate in a cryogen-free dilution refrigerator with a base temperature around 10\,mK. The sample houses eight resonators and eight qubits to provide redundancy against fabrication yield issues related to individual resonators or qubits and to provide a choice of frequency of the emitted microwave photon. The feedline was bonded to the input and output coaxial lines as shown in figure \ref{measurement_setup}. The microwave signal from the room temperature source was attenuated at different temperatures inside the cryostat to suppress the black-body radiation that would otherwise reach the sample. On the output side, the chip was protected from the amplifier input noise and black-body radiation by two isolators mounted at the base temperature. The outgoing signal was amplified by 39\,dB at 3\,K before it was amplified further by 40\,dB at room temperature. The qubit level spacing was controlled through the adjustment of $E_J$ by an external magnetic field threading the qubit SQUID loop, which was generated both by a small superconducting coil placed under the chip holder, and by on-chip flux bias lines capable of producing a magnetic field local to each qubit.

\begin{figure}
\begin{center}
\subfigure[\label{transmission_dip}]{\includegraphics[height=2in]{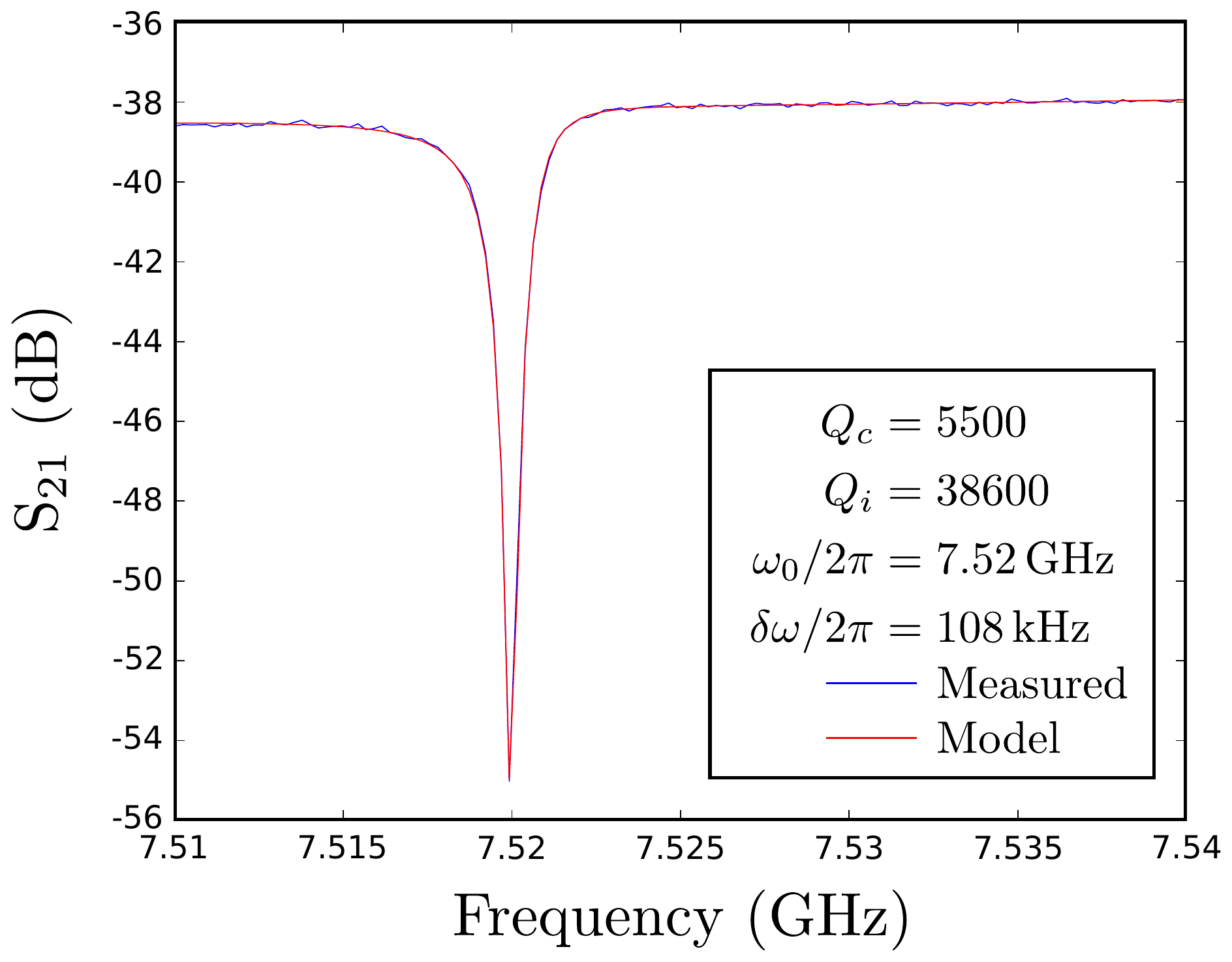}}
\goodgap
\subfigure[\label{chi_plot}]{\includegraphics[height=2in]{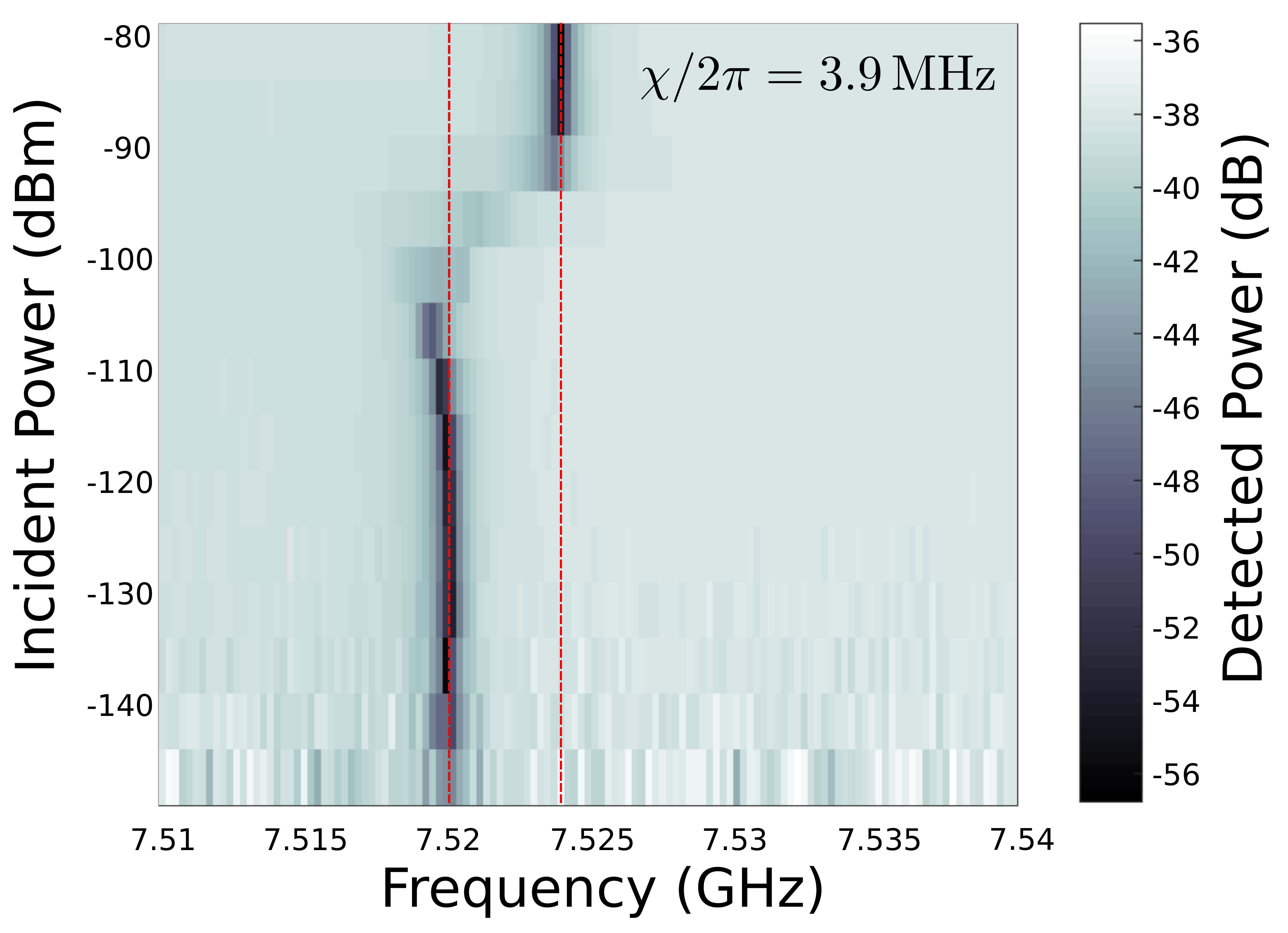}}
\end{center}
\caption{\label{transmission}Continuous wave spectroscopy of the $\lambda/4$ resonator at 7.5\,GHz in the single photon regime, with the qubit tuned to the `sweet-spot' of $\nu_{01}=8.501\,\mathrm{GHz}$. (a) Transmission past the resonator at 7.52\,GHz and (b) resonance frequency as a function of probe power, showing the dispersive shift $\chi$ for a detuning of 990\,MHz.}
\end{figure}

Initial characterisation of the sample was performed by measuring the feedline transmission $S_{21}$ around 6.8 - 7.8\,GHz, which identified the $\lambda/4$ resonators. We chose to focus the study on the resonator at 7.5\,GHz. The transmission past this resonator at an incident power in the single photon regime is shown in figure~\ref{transmission_dip} and corresponds to an external quality factor $Q_c \simeq 5500$, or a resonator decay rate $\kappa / 2 \pi \simeq 1.4\,\mathrm{MHz}$. The resonance lineshape is well-fit using the method of \cite{Khalil2012} by assuming the resonator is overcoupled to the feedline, with an internal quality factor $Q_i \simeq 38600$.

\subsection{$\chi$ measurement}

The presence of the qubit was confirmed by the dispersive shift of the resonator frequency due to the interaction of the resonator and the nonlinear element, which was measured by single tone spectroscopy, sweeping the frequency and power of a probe tone in a continuous wave experiment, and measuring transmission through the feedline with a vector network analyser. As shown in figure \ref{chi_plot}, at input powers below $-110\,$dBm the qubit is in the groundstate \ket{0} and the resonator experiences a frequency shift $- \chi/ 2 \pi$ from the bare resonator frequency. At input powers above $\gtrsim -90\,$dBm, the resonator becomes populated with many photons and the resonance frequency returns to that of the bare resonator \cite{Gambetta2006,Bishop2010}. With the qubit-resonator detuning $\Delta_0 = 990\,$MHz, we observed a dispersive shift $\chi / 2 \pi \simeq 3.9\,$MHz.

\subsection{Qubit-resonator coupling: `$g$' measurement}

\begin{figure}
\begin{center}
\includegraphics[width=0.6\textwidth]{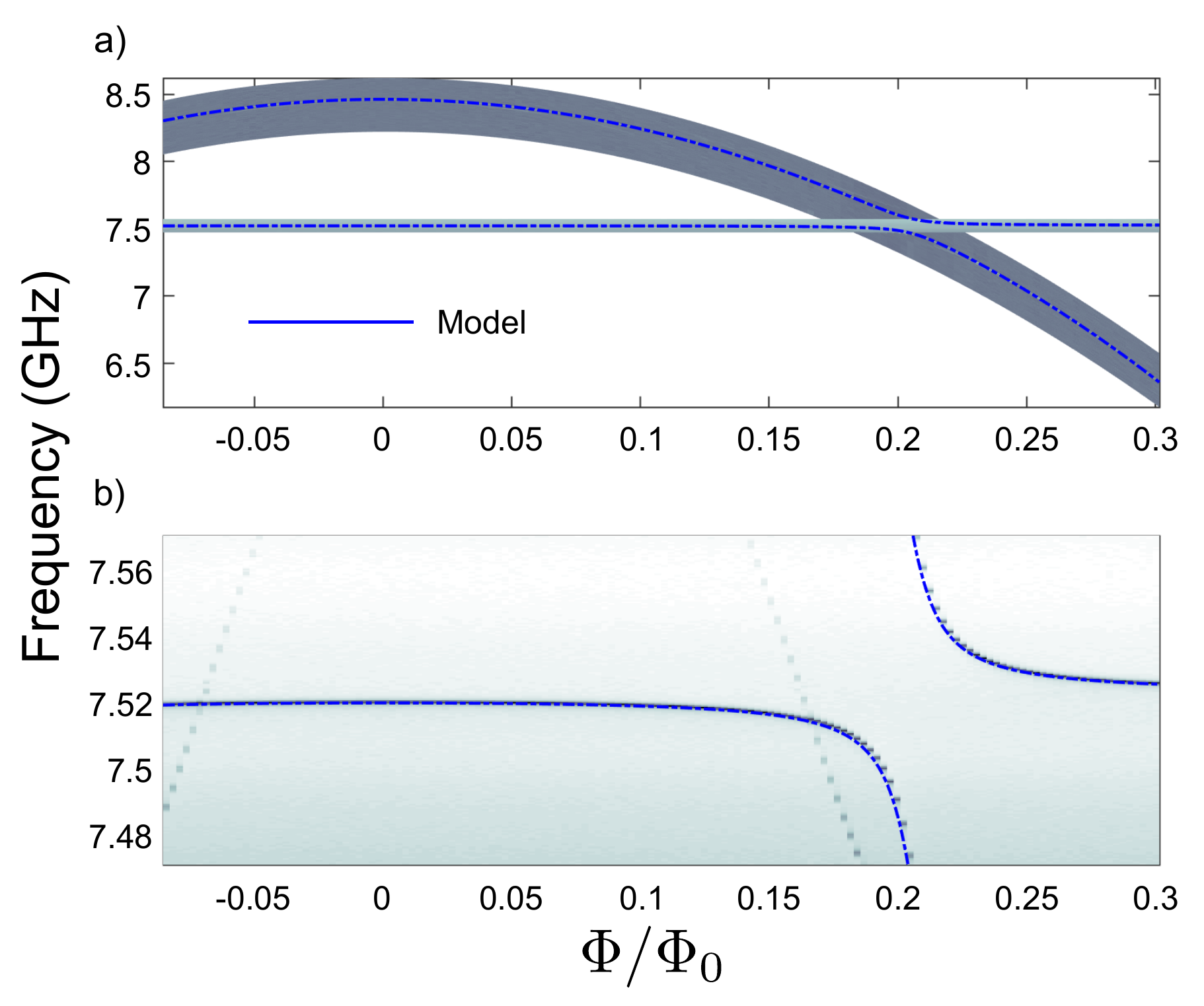}
\end{center}
\caption{\label{anticrossing_figure} (a) Measured transition frequencies of resonator and qubit states determined by monitoring the resonator dispersive shift, as a function of applied flux bias $\Phi$. (b) Close-up of the resonator-qubit hybridisation point, showing the anti-crossing of the resonator and qubit states. (Model of the hybridised resonator-qubit states shown overlaid.}
\end{figure}

Once the presence of the qubit was confirmed, the hybridisation between the resonator and qubit states was investigated by flux tuning the Josephson energy $E_J$ and hence the resonator-qubit detuning $\Delta_0$. The effective Josephson energy of our device follows:

\begin{equation}
E_J = E_J^{\mathrm{max}} \left| \cos \left( \frac{\pi \Phi}{\Phi_0} \right) \right|
\end{equation}

\noindent where $\Phi$ is the magnetic flux through the SQUID loop, $\Phi_0 = h / 2e$ is the flux quantum, and $E_J^\mathrm{max}$ is the Josephson energy at $\Phi = 0$. To achieve flux tuning of the device, we use both a small solenoid external to the sample and an on-chip flux bias line to manipulate the Josephson energy. We use a two-tone spectroscopy technique to investigate the coupled qubit-resonator system \cite{Bianchetti2009}, measuring transmission past the resonator whilst sending a probe tone to excite the qubit. Through use of this method we are able to trace out the qubit and resonator frequencies as a function of the applied magnetic flux $\Phi$, presented in figure \ref{anticrossing_figure}, showing the anti-crossing of the qubit and resonator states at around $\Phi = 0.23\,\Phi_0$. The data are well fit by assuming a resonator qubit coupling of $g / 2 \pi = 54.3\,\mathrm{MHz}$.

\subsection{Deducing the characteristic qubit energies}

The transition frequency of the qubit $h \nu_{01} = \sqrt{8 E_J E_C } - E_C$ is dominated by the $\sqrt{8 E_J E_C}$ term, and is almost insensitive to the $-E_C$ correction. To determine the charging energy of our device spectroscopically, we make use of the dispersive shift $\chi$ which is related to $E_C$ and $\Delta_0$. We have that:

\begin{equation}
\nu_C = \frac{E_C}{h} = \frac{ \chi \Delta_0^2 }{ g^2 / 2 \pi  + \chi \Delta_0}
\end{equation}

\noindent Using our observed values of $g / 2 \pi = 54.3\,\mathrm{MHz}$, $\Delta_0 = 990\,\mathrm{MHz}$, and $\chi/2 \pi = 3.9 \, \mathrm{MHz}$ we deduce $\nu_C = 561 \, \mathrm{MHz}$, which corresponds closely to the $540\,\mathrm{MHz}$ design value targeted in our COMSOL model. We can then solve for the Josephson energy $E_J$ as:

\begin{equation}
E_J = h \nu_J = \frac{(h \nu_{01} + E_C)}{8 E_C}
\end{equation}

\noindent yielding $E_J^{\mathrm{max}} = 76\,\mu\mathrm{eV}$ or equivalently $\nu_J = 18.3 \, \mathrm{GHz}$.

\subsection{Qubit characterisation}
\subsubsection{Rabi oscillations}

\begin{figure}
\begin{center}
\subfigure[\label{chevron}]{\includegraphics[width=0.45\textwidth]{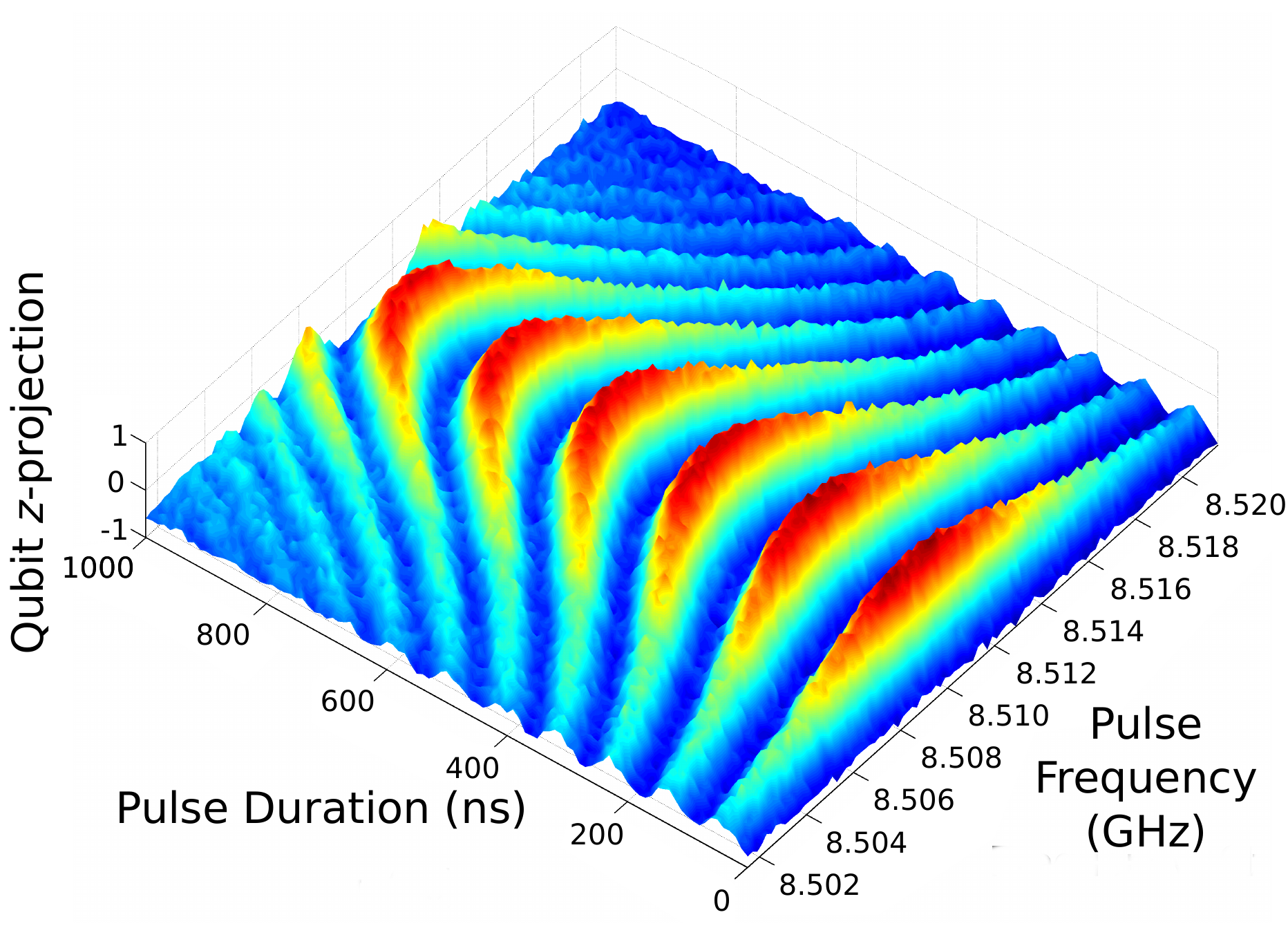}}
\goodgap
\subfigure[\label{chevron_fft}]{\includegraphics[width=0.45\textwidth]{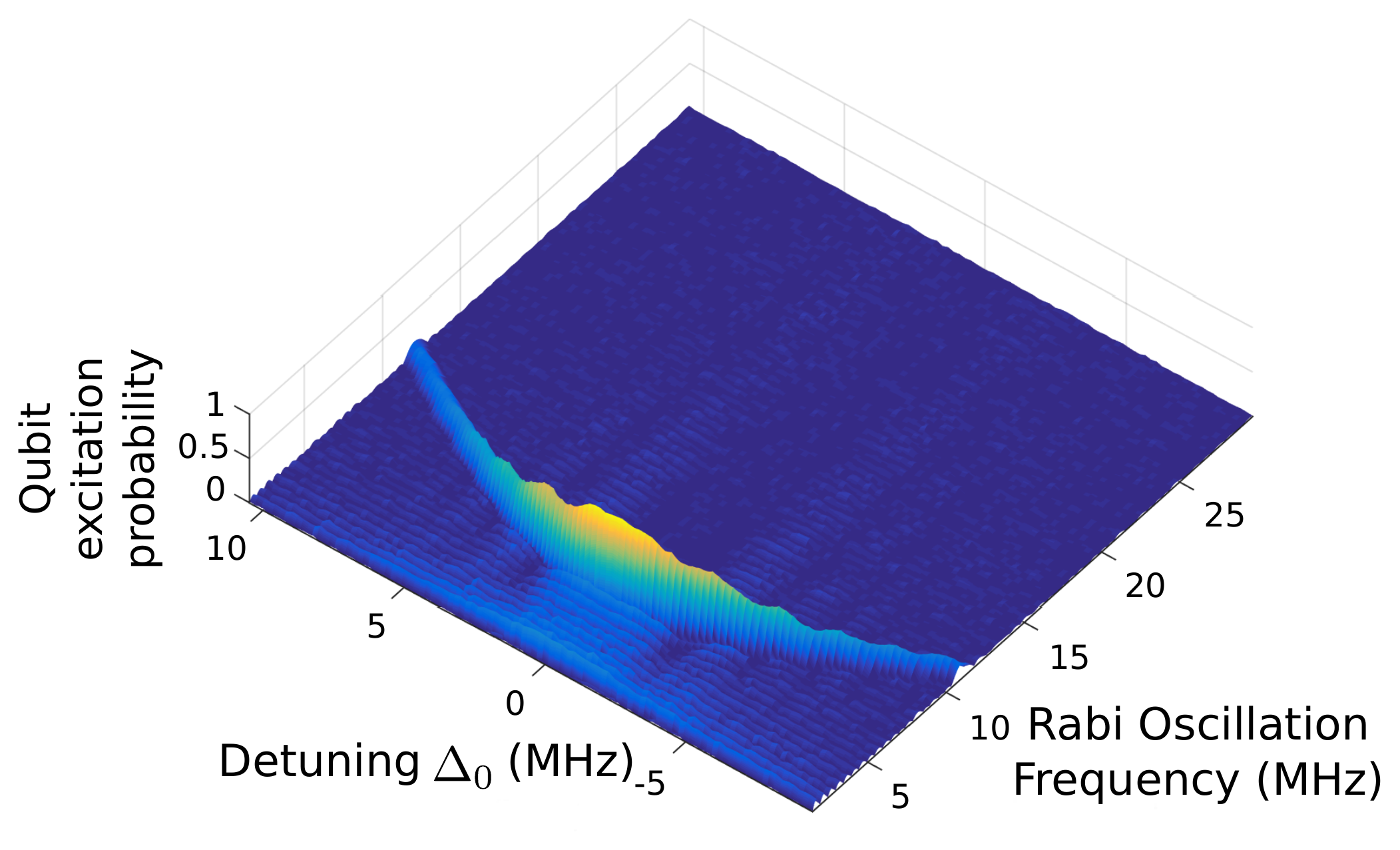}}
\end{center}
\caption{\label{chevron_caption} a) Rabi oscillations recorded as a function of pulse frequency and pulse duration. b) Fourier transform of Rabi oscillation plot against pulse.}
\end{figure}

To characterise the qubit we measured coherent Rabi oscillations around the qubit maximal gap where the energy bands are first order insensitive to the magnetic field (the ``sweet spot''), by sweeping a variable length pulse from 8.502 to 8.520\,GHz, with durations from 0 to 1000\,ns. We inferred the qubit state by measuring the dispersive shift of the resonator through $\pm \chi / 2 \pi$ \cite{Bianchetti2009}. The Rabi oscillations are presented in figure \ref{chevron} and Fourier transformed in figure \ref{chevron_fft}. The oscillations exhibit a typical pattern \cite{Hofheinz2009}, with the lowest oscillation frequency, $\Omega = 6.17\,$MHz, at the drive frequency of 8.512\,GHz when the drive frequency matches the qubit gap. At zero detuning the probability for the qubit to be in the excited state is close to unity. When the drive frequency is detuned by $\Delta_0$ to lower or higher frequency from the qubit sweet spot, the Rabi oscillation frequency increases as $\sqrt{\Omega^2 + \Delta_0^2}$. Also, as expected, when the drive frequency is detuned from the sweet spot the probability of the excited state decreases as $\left. \Omega^2 \middle/ ( \Omega^2 + \Delta_0^2 ) \right.$ and also the qubit decoheres faster. From this data, we found that for our microwave amplitude at zero detuning, it takes about 80\,ns to transfer the qubit from the ground to excited state. The pulse that performs this excitation is called the $\pi$-pulse.


\subsubsection{$T_1$ and $T_2$ measurements}

\begin{figure}
\begin{center}
\subfigure[\label{qubit_characterisation_t1}]{\includegraphics[width=0.45\textwidth]{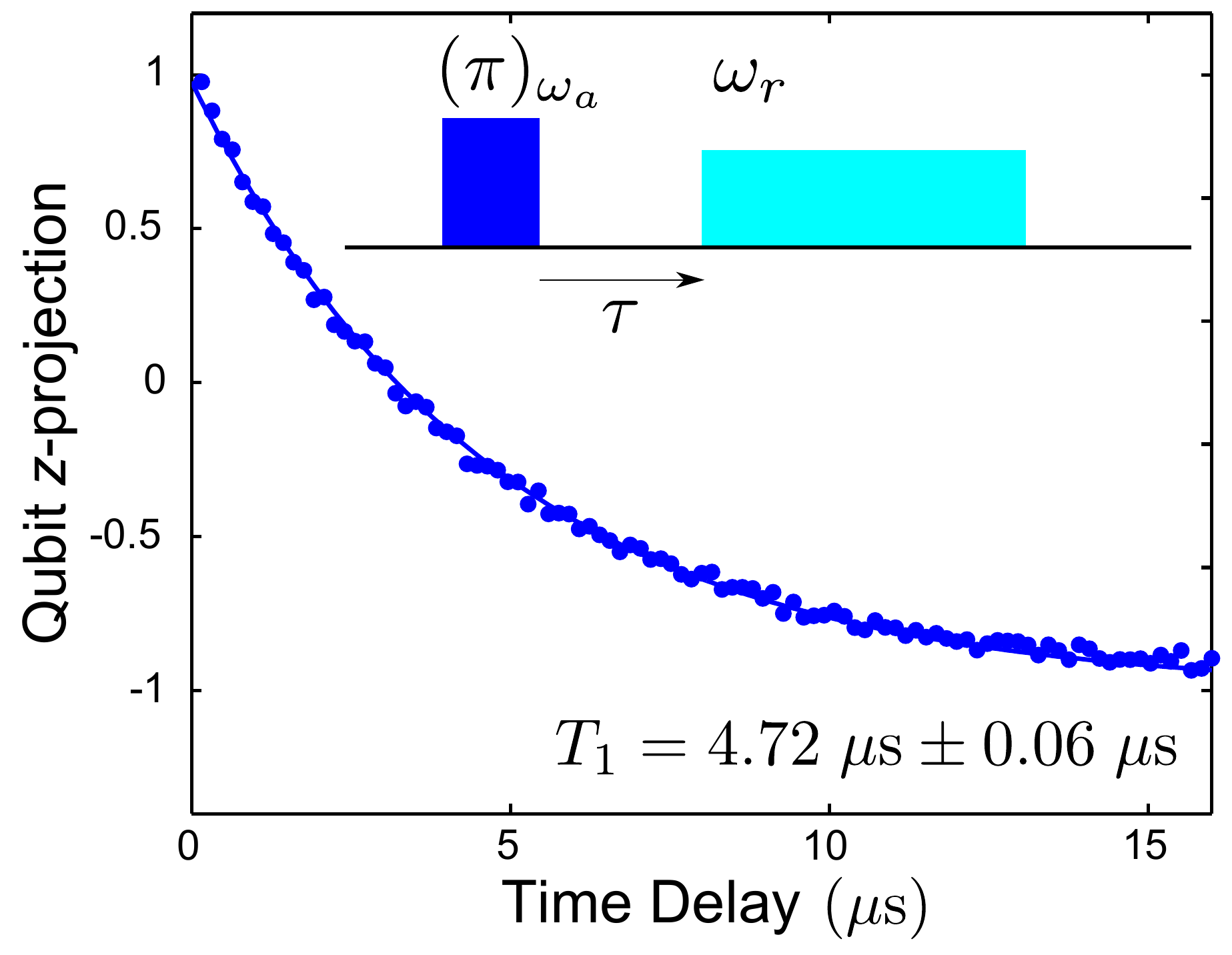}} \\
\subfigure[\label{qubit_characterisation_t2_ramsey}]{\includegraphics[width=0.45\textwidth]{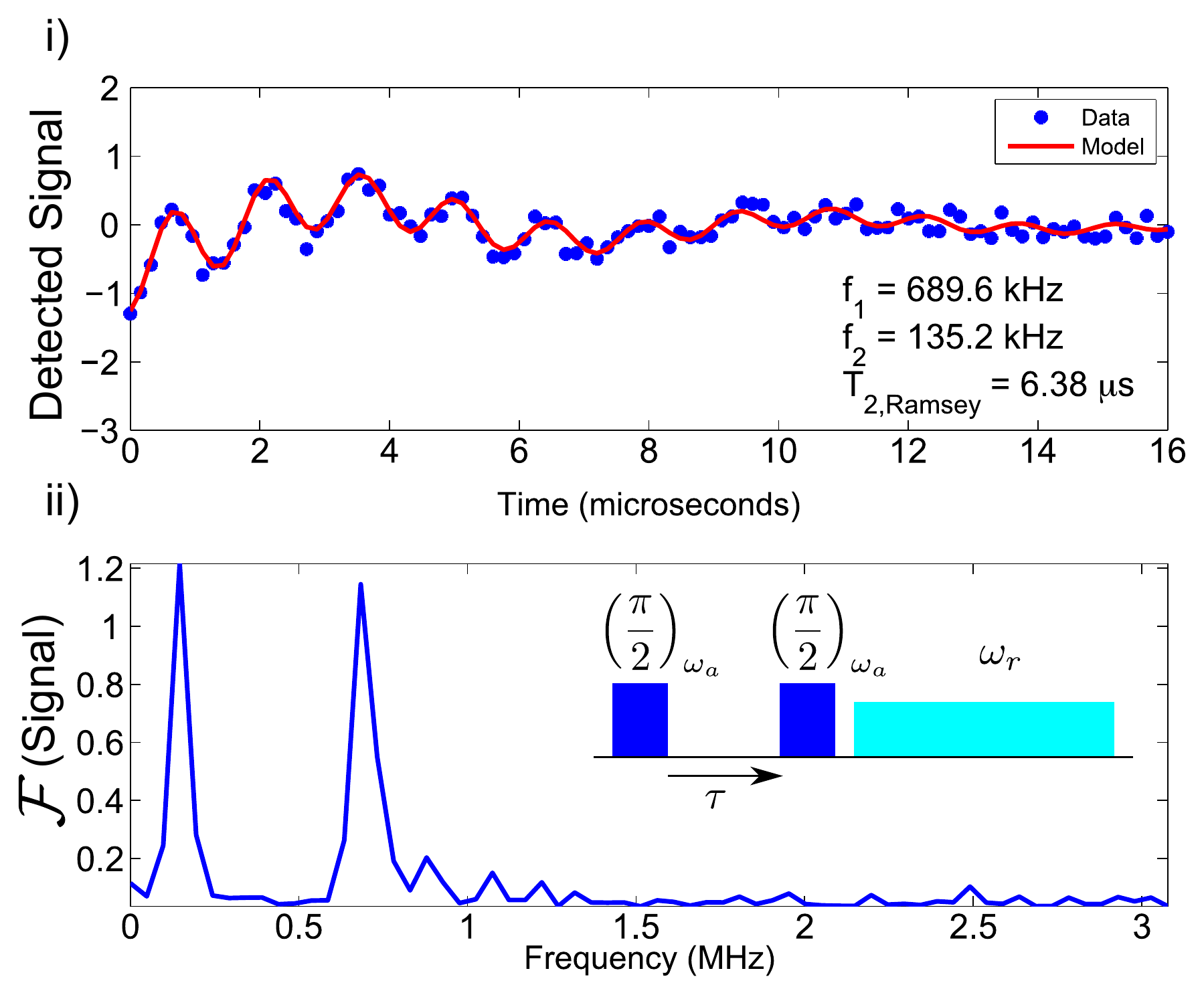}}
\goodgap
\subfigure[\label{qubit_characterisation_t2_echo}]{\includegraphics[width=0.45\textwidth]{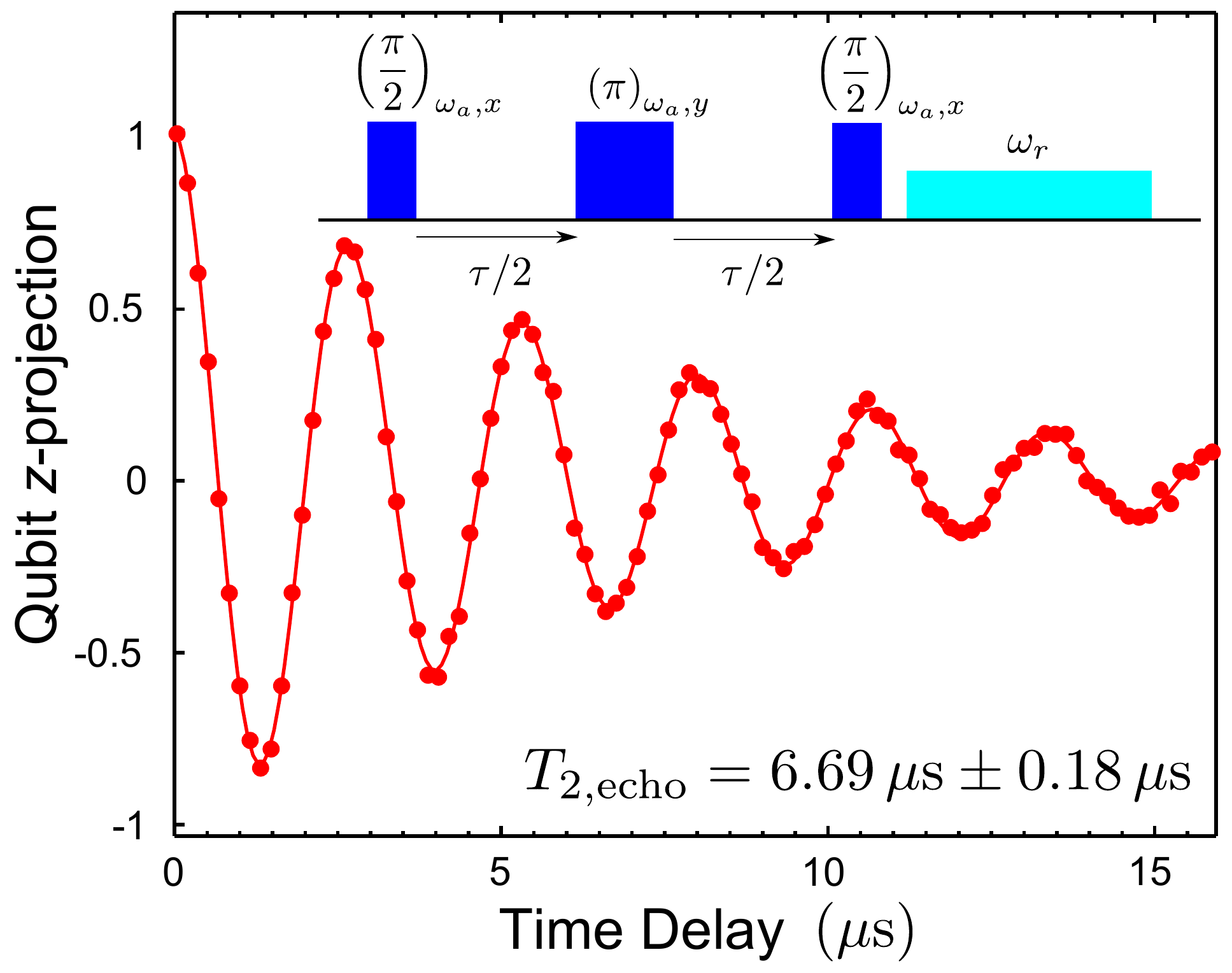}}
\end{center}
\caption{\label{qubit_characterisation} The energy and phase relaxation times of the qubit measured at the `sweet spot' $\nu_{01} \simeq 8.5 \, \mathrm{GHz}$ Insets show the pulse sequence used to record the relaxation rates. (a) Relaxation of the qubit. The fit to the data gives dephasing time $T_1 = 4.72 \pm 0.06\,\mu$s. (b) i. Ramsey fringe measurement of the qubit reveals two oscillating components corresponding to the two different charge parity states of the qubit arising from quasiparticle tunnelling. ii. Fourier transform of the Ramsey fringes. The fit to the data gives $T_{2,\mathrm{Ramsey}} = 6.38\,\mu\mathrm{s}$. (c) Coherent evolution of the qubit with the echo technique applied. The fit to the data gives dephasing time $T_{2,\mathrm{echo}} = 6.69 \pm 0.18\,\mu$s.
}
\end{figure}

To characterise the coherence properties of the transmon qubit, we measured the qubit energy relaxation time $T_1$ by exciting the qubit from the ground to first excited state using a $\pi$ pulse, and varying the delay time $\tau$ between the control and readout pulses as indicated in the inset of figure \ref{qubit_characterisation_t1}. The observed decay of the excited qubit state is described by an exponential relaxation, giving $T_1 = 4.72 \pm 0.06\,\mu $s, which compares favourably with the transmon relaxation times measured in earlier experiments \cite{Houck2008,Schreier2008,Hofheinz2009,Chow2012}.

The qubit dephasing time was first measured using the Ramsey fringe visibility technique of two $\pi/2$ pulses separated by a variable delay $\tau$, to include the influence of low frequency noise that may adversely affect the photon source during operation. We found $T_{2,\mathrm{Ramsey}} = 6.38 \,\mu\mathrm{s}$ presented in figure \ref{qubit_characterisation_t2_ramsey}i. Fourier transforming the Ramsey fringe data (figure  \ref{qubit_characterisation_t2_ramsey}ii) reveals two oscillating components of equal magnitude, separated in frequency by 554\,kHz. We interpret these two components as arising from quasiparticle tunnelling events that take place on a characteristic timescale slow compared to each shot of the experiment, but fast compared to the $10^4 \sim 10^5$ realisations that are averaged to provide the detected signal \cite{Riste2013}.

\begin{figure}
	\begin{center}
		\includegraphics[width=0.45\textwidth]{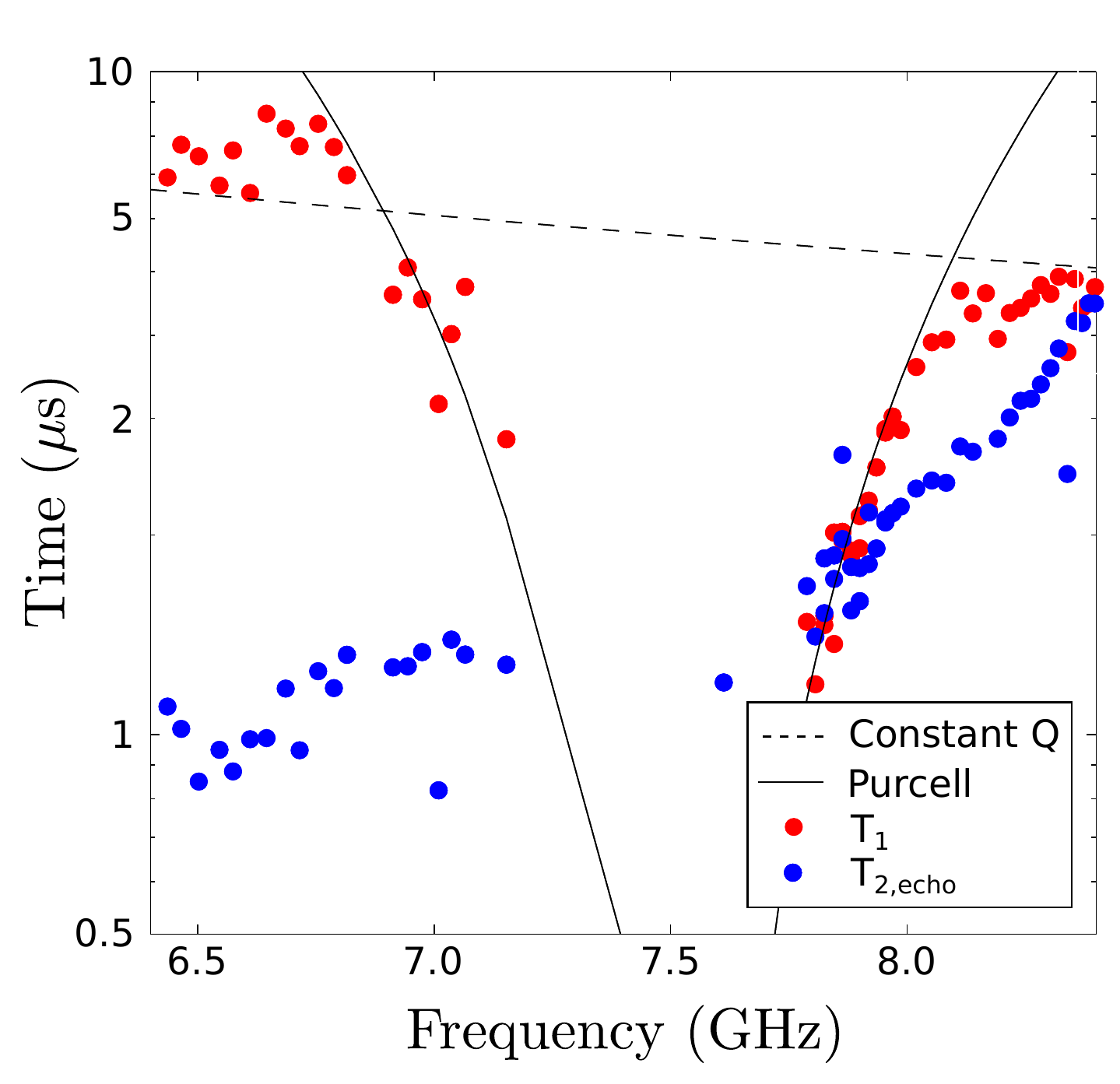}
	\end{center}
	\caption{\label{purcell}$T_1$ and $T_{2,echo}$ measured as a function of qubit transition frequency, showing an enhanced relaxation rate as the qubit transition frequency approaches that of the resonator.}
\end{figure}

The echo technique was then used to eliminate this low-frequency noise, using a $(\pi/2) - \tau/2 - (\pi) - \tau/2 - (\pi/2)$ pulse sequence and recording the resonator dispersive shift to infer the qubit state. We found a value of $T_{2,\mathrm{echo}} = 6.69 \pm 0.18\, \mu\mathrm{s}$ presented in figure \ref{qubit_characterisation_t2_echo}. Schematics of the pulse sequences used are shown in the insets of figure \ref{qubit_characterisation}. The $T_1$ and $T_{2,\mathrm{echo}}$ lifetimes of the qubit were then measured as a function of the qubit level spacing $\nu_{01}$, which was swept by varying the magnetic field and hence $E_J$. At the $\Phi = 0$ `sweet-spot' where the Josephson energy is first-order insensitive to magnetic field, $T_1$ and $T_2$ take their maximum values. Once the qubit-resonator detuning $\Delta_0$ begins to fall towards zero, the $T_1$ value is drastically reduced due to the Purcell effect, corresponding to relaxation of the qubit by microwave photon emission from the resonator. $T_1$ and $T_{2,\mathrm{echo}}$ as a function of $\nu_{01}$ are plot in figure \ref{purcell}.

\section{Discussion}

We now turn to the use of this device as a single microwave photon source. There are two different regimes in which a qubit coupled to a cavity can be used to generate single microwave photons: The qubit level spacing $\nu_{01}$ can be static \cite{Houck2007} or dynamically tuned to exchange photons with the resonator \cite{Bozyigit2011}. While our device was designed for dynamic tuning, we will start by discussing the use of the simpler static case. Here the qubit-resonator detuning is fixed, the qubit is excited by a pulse from the flux tuning line, and the qubit emits a photon into the resonator which in turn releases the photon to the transmission-line. The efficiency of such a source is:

\begin{equation}
\eta_\mathrm{static} = \epsilon \left. \frac{f_0}{Q_c} \left( \frac{g}{2 \pi \Delta_0} \right)^2 \middle/ \left( \frac{1}{T_1} + \frac{2}{T_2} \right) \right.,
\end{equation}

\noindent where $0 \le \epsilon \le 1$ is the polarisation of the qubit after applying the $\pi$-pulse. We estimate $\epsilon \approx 1 - \tau_{\pi} / T_1 = 98.3 \%$, where $\tau_\pi = 80\,\mathrm{ns}$ is the duration of our $\pi$-pulse. For the parameters of our device we obtain $\eta_\mathrm{static}=0.9\%$ as the probability that the photon will leave the resonator with the qubit at the sweet-spot. Tuning the qubit towards the strong Purcell limit we can approach $\eta_\mathrm{static}=1.8\%$ based on the data in figure 8. To reach a much higher source efficiency with static qubit tuning, the resonator must be coupled strongly to the transmission line. The current coupling $Q_c = 5500$ is instead close to the optimal value for a dynamic protocol to generate single photons with much higher efficiency.

The dynamic protocol calls for the qubit to first be excited by a $\pi$-pulse whilst detuned from the $\lambda/4$ resonator, followed by use of the flux tuning line to bring the qubit into resonance with the $\lambda/4$ structure for a duration $1/(2g) = 58\,\mathrm{ns}$, effectively swapping the excitation from the qubit to the resonator. In our experiment the flux tuning line used to modulate $E_J$ has a bandwidth of 1\,GHz, meaning that we can bring the qubit and resonator into resonance with one another in $\sim$1\,ns. The time taken to tune the qubit is therefore negligible compared to the $\pi$-pulse duration (80\,ns) and vacuum Rabi swap (58\,ns).

Once the excitation is transferred to the $\lambda/4$ resonator, the photon can either be absorbed by loss mechanisms in the resonator, or emitted into the feedline. The efficiency of the source under these operating conditions is therefore:

\begin{equation}
\eta_\mathrm{dynamic} = \frac{Q_i}{Q_i + Q_c} \exp \left( - \frac{\tau_\pi}{T_1} - \frac{1}{2 g T_1} \right),
\end{equation}

\noindent this expression being valid in the fast driving and strong coupling limit where $g T_1 >> 1$ and $2 \pi \,\Omega T_1 >> 1$. For our device, we find that $\eta_\mathrm{dynamic} \approx 85 \%$.

The efficiency of the photon source when dynamic tuning is employed is limited principally by the loss mechanisms of the $\lambda/4$ resonator into which the photon is emitted. Our photon source efficiency could be improved by increasing $Q_i$. The best internal quality factor reported for 2D superconducting resonators at low excitation power is $Q_i \sim 2 \times 10^6$ \cite{Megrant2012}.

The source efficiency can also be improved through lowering $Q_c$, however $Q_c$ cannot be lowered arbitrarily, as the Purcell effect will begin to shorten $T_1$ even when the resonator and qubit are detuned during the excitation phase of the dynamic protocol. There is then an optimal value of $Q_c$ for a given qubit $T_1$ and resonator frequency $\omega_r / 2 \pi$. We have that:

\begin{equation}
Q_c^\mathrm{optimal} = 2 \pi \omega_r T_1 \left( \frac{g}{2 \pi \Delta_0} \right)^2
\end{equation}

The best $T_1$ times reported in the literature for superconducting qubits are a factor of 15 higher than the device realised in this paper \cite{Paik2011,Rigetti2012}. If we could build a device whose $T_1 \simeq 70\,\mu\mathrm{s}$ we would be able to decrease $Q_c$ to 3500. If we were also able to make use of higher quality resonators, then we could hope to achieve a source efficiency $\eta_\mathrm{opt} > 99.5\%$.

\section{Conclusions}

We have fabricated and characterised a superconducting qubit multiplexing circuit that appears to be promising for single microwave photon generation. We note that the resonators' frequencies and quality factors are close to the designed ones. The qubit parameters, such as the maximal energy gap, which exceeds the resonator frequency by about 1$\,$GHz, are also close to the desired ones. Finally, the qubit-resonator and resonator-feedline couplings appeared to be consistent with expectations. The measured qubit coherence times and resonator intrinsic quality factors are consistent with those reported by other groups, but can be further improved to produce a single microwave photon source with an efficiency approaching 100\%.

\section*{Acknowledgements}

This work was carried out within the project EXL03 MICROPHOTON of the European Metrology Research Programme (EMRP). EMRP is jointly funded by the EMRP participating countries within EURAMET and the European Union. YuAP acknowledges partial support by the Royal Society (Grant WM110105). JS, OPS and JP acknowledge support from Academy of Finland grants 284594 and 272218.

{\it Authour contributions:} The device design was discussed by all authors. Iterations of device fabrication and measurement were performed by REG and SEdG using facilities at Lancaster University and Chalmers University of Technology, and by JS, REG, and OPS using facilities at Aalto University. REG, SEdG and YuAP wrote the paper with comments from all authors.

\section*{References}

\bibliographystyle{jphysicsB}

\end{document}